\documentclass[aps,prd,twocolumn,showpacs,superscriptaddress,nofootinbib,tightenlines,floatfix]{revtex4-2}

\usepackage{hyperref}
\usepackage{physics}
\usepackage{amsmath,amssymb}
\usepackage{graphicx,multirow,tabularx}
\usepackage{davitex}
\hypersetup{
    colorlinks=true,     
    linkcolor=blue,      
    citecolor=blue,      
    filecolor=blue,      
    urlcolor=blue,        
    linktoc=page
}

\usepackage{color}
\usepackage{xcolor}

\newcommand{\chimax}{{\rm max}\left(\chi_{{\rm MSE}}\right)}
\newcommand{\MSE}{{\rm MSE}}
\newcommand{\Tlow}{T_{\rm lower}}
\newcommand{\Thi}{T_{\rm upper}}

\newcommand{\houston}{University of Houston, Houston, Texas 77204, USA}

\newcommand{\TcRes}{4.5128(58)}
\newcommand{\nuRes}{0.63(27)}

\begin{document}

\title{Detecting the 3D Ising model phase transition with a ground-state-trained autoencoder}

\author{Ahmed Abuali}\email{amabuali@uh.edu}\email{ahmed.abuali1975@gmail.com}
\affiliation{Physics Department, \houston}

\author{David A. Clarke}\email{clarke.davida@gmail.com}
\affiliation{Fakult\"at f\"ur Physik, Universit\"at Bielefeld, D-33615 Bielefeld, Germany}

\author{Morten Hjorth-Jensen}\email{mhjensen@uio.no}
\affiliation{Department of Physics and Center for Computing in Science Education, University of Oslo, N-0316 Oslo, Norway}

\author{Ioannis Konstantinidis}
\affiliation{Computer Science Department, \houston}

\author{Claudia Ratti}
\affiliation{Physics Department, \houston}

\author{Jianyi Yang}
\affiliation{Computer Science Department, \houston}

\date{\today}

\begin{abstract}
We develop a one-class, deep-learning framework to detect the phase transition and 
recover critical behavior of the 3D Ising model. A 3D convolutional neural network autoencoder (CAE) is
trained on ground-state configurations only, without prior knowledge of the
critical temperature, the Hamiltonian, or the order parameter. After training,
the model is applied to Monte Carlo configurations across a wide temperature range and
different lattice sizes. The mean-square reconstruction error is shown to be sensitive to the transition.
Finite-size scaling of the peak location for the reconstruction
error susceptibility yields the critical temperature $\Tc=\TcRes$ and the 
correlation-length critical exponent $\nu=\nuRes$, consistent with results from
the literature. Our results show that a one-class CAE, trained on 
zero-temperature configurations only, can recover nontrivial critical behavior 
of the 3D Ising model.
\end{abstract}

\maketitle

\section{Introduction}\label{sec:intro}

In recent years, machine-learning (ML) techniques have emerged as powerful tools for discovering structure in many-body systems directly from microscopic data. Supervised neural networks have been successfully trained to distinguish phases when supplied with labeled data~\cite{Carrasquilla:2016oun,%
Li:2017xaz,%
Cossu:2018pxj,%
ShibaFunai:2018aaw,%
Tan:2019eih,%
Bachtis:2020dmf,%
Walker:2020hiq,%
Li:2025rbf,%
Abuali:2025vmn},
while unsupervised techniques, including but not limited to principal component analysis (PCA), 
clustering algorithms, and autoencoders reveal that the enormous space of spin configurations actually 
contains structures that correspond to physical 
ordering~\cite{Wang:2016nmn,%
Hu:2017wey,%
Wetzel:2017olt,%
Rodriguez-Nieva:2018cbl,%
Alexandrou:2019hgt,%
Han:2022mhy,%
civitcioglu_2025,%
jang_unsupervised_2025}. 
Generically speaking, supervised ML models are trained on many temperatures
above and below the transition. In
Refs.~\cite{Li:2017xaz,Tan:2019eih,Li:2025rbf,Abuali:2025vmn}, minimal training set
models, i.e. models training on one or two out-of-distribution temperatures, 
were shown to be capable of recognizing
phases and, in some cases, extracting critical parameters. Choosing one or two
sets with unambiguous phases is a strategy that allows one to train even in 
cases where labeling is challenging, for example in systems with crossovers, 
and it has the computational advantage of significantly reducing the training time.

Given the analytic tractability of the 2D Ising model,
studies showcasing the capabilities of ML to detect its phase transitions are
plentiful; in addition to some of the above studies, one finds numerous
examples in e.g. Ref.~\cite{Aarts:2023vsf} and references therein. Such studies 
for the 3D Ising model, meanwhile, remain comparatively 
sparse~\cite{kim_smallest_2018,%
chng_unsupervised_2018,%
zhang_few-shot_2019,%
Li:2021yet,%
Guo:2023dpp,%
panda_non-parametric_2023,%
jang_unsupervised_2025}.

In this paper we extend  ML studies of the 3D Ising model by applying a 
minimally trained network to extract critical
parameters of the system. In particular, we adopt a similar approach
as our previous study of the 2D Ising model~\cite{Abuali:2025vmn}, using an 
architecture with convolutional layers trained on only $T=0$ configurations. 
Moving from two to three spatial dimensions considerably increases the complexity
of the input; therefore, instead of a convolutional neural network binary classifier, we 
employ a convolutional neural network autoencoder (CAE) to find simplified representations
of the configurations. We find that the CAE output responds to changes in
phase and is capable of accurately extracting $\Tc$ and $\nu$.

The outline of this paper is as follows. We start in \secref{sec:ising} 
with a brief summary of the 3D Ising model, along with results from
statistical physics studies that will be used to benchmark our CAE.
In \secref{sec:setup}, we describe our Monte
Carlo data, ML architecture, and training strategy. The output
of the CAE is presented in \secref{sec:results}, showing this approach can
successfully extract $\Tc$ and $\nu$. We close with \secref{sec:summary}, 
where we discuss our findings and the outlook.
\section{3D Ising model and finite-size scaling}\label{sec:ising}
The 3D Ising model consists of spins $\sigma_i = \pm 1$ residing on 
the sites of a lattice of volume $V=L^3$ with Hamiltonian
\begin{equation}
    H = -J \sum_{\langle i\,j \rangle} \sigma_i \sigma_j,
\end{equation}
where the sum runs over nearest-neighbor pairs and $J$ is the interaction strength.
For the remainder of this paper, we work in units
$J=k_B=1$. The order parameter is the absolute value of the magnetization $|m|$, where
\begin{equation}
    m = \frac{1}{V}\sum_{i} \sigma_i.
\end{equation}
As the system approaches the critical temperature $\Tc$, the magnetic susceptibility
\begin{equation}
    \chi = \frac{V}{T}\left(\langle |m|^2\rangle - \langle |m|\rangle^2\right)
\end{equation}
diverges. For finite $L$, the divergence in
$\chi$ softens to a pronounced peak, whose location in $T$ can be used to define a
pseudocritical temperature $\Tc(L)$. This pseudocritical temperature approaches $\Tc$
according to the critical exponent $\nu$ as
\begin{equation}\label{eq:FSS}
|\Tc(L)-\Tc|\sim L^{-1/\nu}.
\end{equation}

Although the 2D Ising model admits an exact solution~\cite{Onsager:1943jn}, no closed-form 
solution exists in 3D. A common strategy to glean quantitative understanding of the 3D
Ising model is to use Markov chain Monte Carlo (MCMC) simulations. MCMC studies
have delivered robust results for the critical parameters over the past several decades.
Taking a weighted average of many (mostly MCMC) literature
results~\cite{Pawley:1984et,Barber:1983mp,livet_cluster_1991,ito_monte_1991,blote_finite-size_1993,LANDAU199441,Blote:1995zik,Talapov:1996yh,gupta_critical_1996,Kim:1996zz,Blote:1999vw,Hasenbusch:2010hkh} 
yields $\Tc = 4.511535(31)$; meanwhile, a weighted average for $\nu$
\cite{LANDAU199441,Blote:1995zik,gupta_critical_1996,Hasenbusch:1996eu,Hasenbusch:1998gh,Ballesteros:1998zm,Blote:1999vw,El-Showk:2014dwa,Kos:2016ysd} 
delivers $0.62999(34)$. 

\section{ML model and MCMC set up}\label{sec:setup}

In our previous study~\cite{Abuali:2025vmn}, we were interested in binary
classification using training approaches that could be employed also when 
ensemble labels are not readily available, for example when one does not
know the location of $\Tc$ {\it a priori}.
This motivated us to adopt a training regimen where
the network is only shown classes at $T=0$ and $T=\infty$, where
the only required knowledge is the form of maximally ordered and 
maximally disordered configurations.
Previous investigations of minimally trained networks have shown that only
a single training class, in particular the ground state, is sufficient for
supervised models to detect phase changes in statistical physics
systems~\cite{Li:2017xaz,Tan:2019eih}. Moreover, while training sets
at $T=0$ are exact, sets at $T=\infty$, realized as configurations with
half up spins and half down spins, are only a very rough approximation,
intended primarily to show the network what a maximally disordered configuration
looks like. With this in mind, and motivated by the one-class studies,
we choose to employ a one-class training regimen here. 

The network we used for our 2D Ising study had one convolutional layer, because
another 2D Ising study found that the same architecture was well optimized for
the 2D Ising model when trained on multiple temperatures above and below
$\Tc$~\cite{Bachtis:2020dmf}. While we found that this architecture performed 
reasonably well for a two-set regimen in our previous study, it cannot be
expected to succeed when applied to a one-set regimen with an additional spatial
dimension. To help tackle the increased complexity when moving to 3D, we wanted
an approach that can find reduced representations of the input
configurations to expose their salient features.
Moreover, using an autoencoder allows us to avoid an explicit label
for our training set; hence the approach taken in this study is unsupervised.

Therefore, we use a one-class 3D CAE, which is trained
exclusively on ground-state configurations,
i.e. configurations at $T=0$. 
The autoencoder consists of an encoder $e$ and a decoder 
$d$, which together define the reconstruction of an input
configuration $C$ as
\begin{equation}
\bar{C}=d\left(e\left(C\right)\right),
\end{equation}
where $\bar{C}$ denotes the reconstructed spin configuration.
The encoder progressively reduces the spatial resolution of the input
configuration using strided 3D convolutions, thereby compressing the spin
configuration into a lower-resolution representation. The decoder then 
reconstructs the configuration by applying a sequence of 3D transpose 
convolutions, which increase the spatial resolution in a learnable manner 
until the original lattice size is recovered. 
Autoencoders are well suited for this task because they learn the statistical
structure of a dataset directly from training sets without requiring labels or 
predefined physical observables. Details of the architecture are
discussed in Appendix~\ref{ap:CAE}.

To quantify the quality of the reconstruction $\bar{C}$, we employ
the mean-square error (MSE) 
\begin{equation}
\MSE\left(\bar{C},C\right)\equiv\frac{1}{V}||\bar{C}-C||_2^2=\frac{1}{V}\sum_i \left(\bar\sigma_i-\sigma_i\right)^2,
\end{equation}
where the sum runs over all sites and $\bar\sigma_i$ is the reconstructed
spin value of $\bar{C}$ at site $i$. Since the training is performed only on ground-state
configurations, the MSE evaluated on $T>0$ configurations
measures how well $C(T)$ can be represented by the structure learned
from the ground-state data. Configurations similar to those seen during
training produce small MSEs, while thermally disordered
configurations produce larger errors.

Spin configurations of size $L^3$ with $50\leq L\leq130$ for the 3D Ising model 
are generated using Markov chain Monte Carlo (MCMC). For each $L$, we probe 84
temperatures in the range $T\in[3.0,6.0]$. 
The temperatures are not evenly
spaced, but rather more dense close to the literature value of $\Tc$. 
Simulations are adapted from the Metropolis
checkerboard algorithm implementation described in
Ref~\cite{jacek_wojtkiewicz_monte_2015}.
For each lattice size, the system is first thermalized using a number of warm-up 
sweeps equal to 20\% of the total number of Monte Carlo sweeps. For each
temperature we generate 3000 configurations; for temperatures near $\Tc$, we
generate 4000 configurations to enhance statistical resolution. After
thermalization, measurements are separated by $L^2$ sweeps. For $L\leq60$,
configurations near $\Tc$ are instead separated by $L^3$ sweeps
to reduce autocorrelation between successive measurements.
This choice reflects the increase of the autocorrelation times with system size, 
particularly near the critical temperature.
All MCMC configurations are evaluated by the 
trained neural network without further preprocessing.
These configurations are used exclusively for evaluation; no $T>0$ data are 
included in the training of the autoencoder. 

The autoencoder is trained on 2000 ground-state $T=0$ configurations, which
consist of 1000 fully spin-up configurations and 1000 fully spin-down configurations. 
Including both ordered states ensures that the training set respects the global 
$\mathbb{Z}_2$ symmetry of the Ising model.
After training, the model is applied to the MCMC configurations discussed above.
Each configuration is represented as a 3D binary field
which serves as the input to the machine-learning model. 
No physical observables, temperature labels, or Hamiltonian parameters 
are provided to the network. In these respects, the CAE is completely physics-agnostic.

Statistical analysis, curve fitting, and plotting was performed using
AnalysisToolbox software~\cite{Clarke:2023sfy}.



\begin{figure*}
    \centering
    \includegraphics[width=0.5\linewidth]{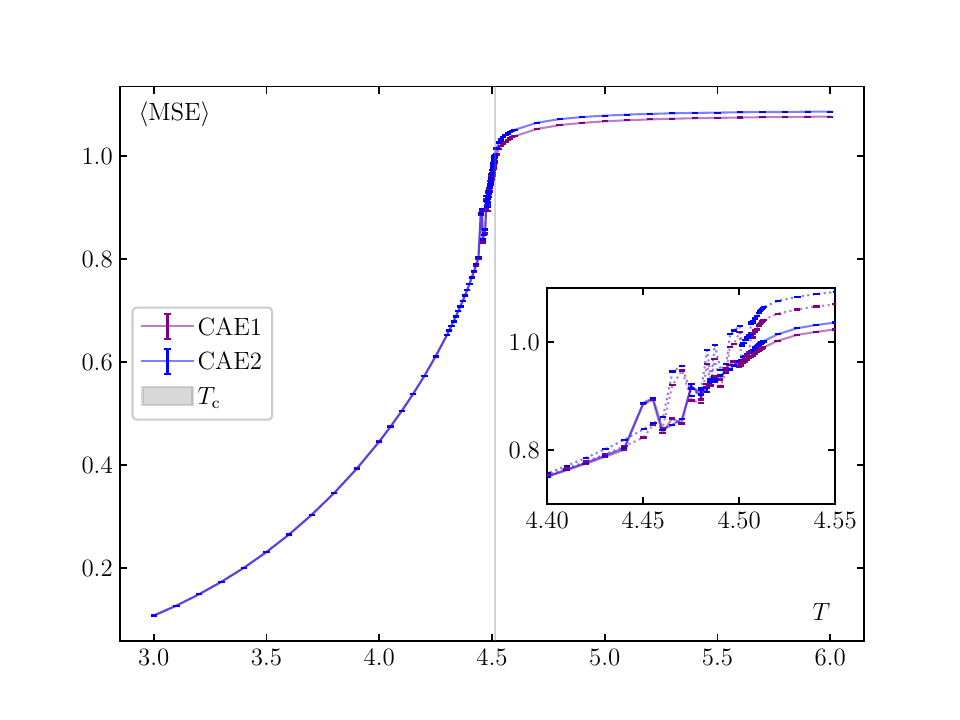}\hspace{-6mm}
    \includegraphics[width=0.5\linewidth]{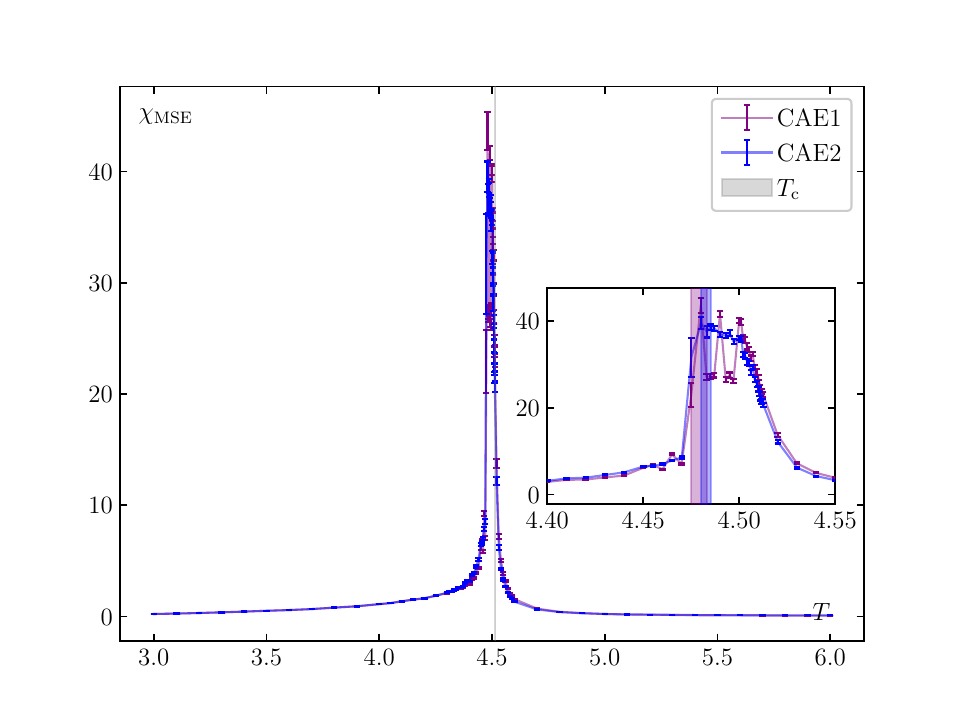}
    \caption{MSE-derived quantities coming from the CAE.
    The vertical, gray band indicates the literature average $\Tc=4.511535(31)$.
    {\it Left}: Ensemble average of the MSE for our smallest ($L=50$) and 
    largest ($L=130$) lattices from two runs of the CAE. Solid lines, drawn to guide
    the eye, indicate $L=50$ while dashed lines indicate $L=130$. {\it Right}:
    $\chi_{\rm MSE}$ for $L=50$ from two runs of the CAE. Vertical bands in
    the inset indicate the extracted $\Tc(50)$ for each run.
    }
    \label{fig:errAndSusc}
\end{figure*}

\section{Results}\label{sec:results}

After applying our CAE to our MCMC data, we estimate $\ev{\MSE}$ and its statistical
uncertainty for each temperature using a standard jackknife with 40 bins.
In \figref{fig:errAndSusc} (left) we show results for $\ev{\MSE}$. To check the stability
of the CAE, we tested 
two independent (i.e. different starting seeds)
CAEs on the same Markov chains, which we call CAE1 and CAE2, indicated in purple and 
blue. We show results from our $L=50$ ensembles (solid lines) and our $L=130$ 
ensembles (dotted lines). The inset shows the region $4.40<T<4.55$. We see 
that $\ev{\MSE}$ generically
increases with temperature, which is the expected behavior, since the CAE is trained
on ground-state configurations, and increasing the temperature drives configurations
to be less ordered, i.e. they bear less resemblance to the ground state. In the
inset, we can see that $\ev{\MSE}$ changes between CAE runs, though this
discrepancy seems to be smaller than the dependence on $L$.
We see also that the
increase of $\ev{\MSE}$ with $T$ is not strictly monotonic: in the region near
$\Tc$, the curves become somewhat jagged, and this non-monotonicity increases 
for $L=130$.  Similar jaggedness has been observed
when using a quantity derived from the MSE in the 2D $J_1$-$J_2$ Ising 
model~\cite{civitcioglu_2025}, so it may also be that this quantity is
inherently quite sensitive to the medium-scale structure of configurations.
Jaggedness near $\Tc$ could be a reflection of the confusion of the neural network
near the boundary between two phases.
Still, we find in all cases a drastic change 
in curvature near $\Tc$, which is suggestive of the $\MSE$ feeling criticality. 

\begin{figure*}
    \centering
    \includegraphics[width=0.5\linewidth]{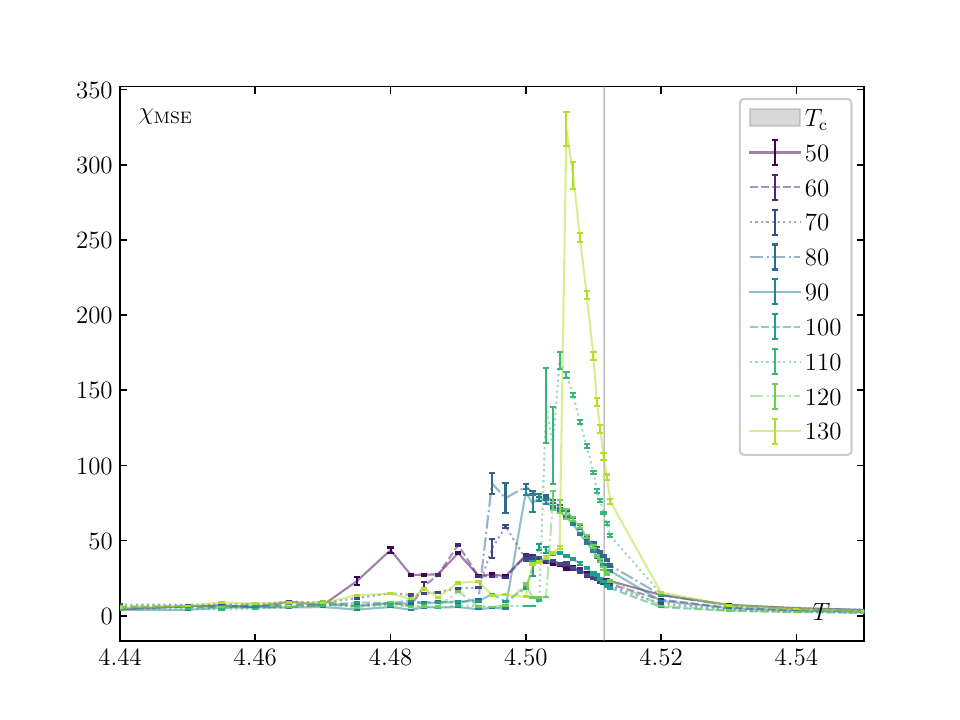}\hspace{-6mm}
    \includegraphics[width=0.5\linewidth]{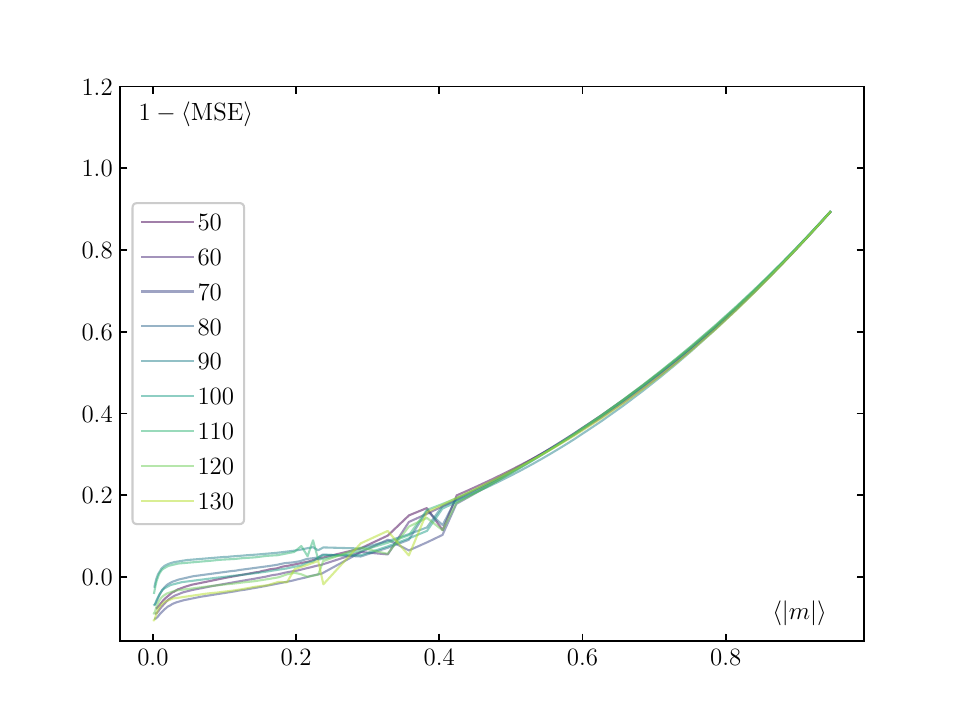}
    \caption{{\it Left}: Dependence of $\chi_{\rm MSE}$ vs $T$ on $L$ for CAE1. The
    vertical, gray band indicates $\Tc$.
    {\it Right}: Correlation of $1-\ev{\MSE}$ with the order parameter for CAE1. 
    Statistical uncertainties are suppressed for visibility.
    }
    \label{fig:errCorrelation}
\end{figure*}

Given the apparent sensitivity of $\ev{\MSE}$ to the phase change, we examine
the susceptibility
\begin{equation}
\chi_{\MSE} = \frac{V}{T}\left(\ev{\MSE^2}-\ev{\MSE}^2\right).
\end{equation}
Results for $\chi_{\MSE}$ at $L=50$ are shown in \figref{fig:errAndSusc} (right). 
We see a sharp peak near $\Tc$ for both CAE1 and CAE2, and in fact, we found
such a peak structure for all lattice sizes. We use the maximum of this
susceptibility, $\chimax$,
to define a pseudocritical temperature $\Tc(L)$; namely
we locate $\chimax$ and use the temperature immediately to its left,
$\Tlow$, and immediately to its right, $\Thi$, estimating
\begin{equation}
\Tc(L)=\frac{1}{2}(\Thi+\Tlow),~~~~~\sigma_{\Tc(L)}=\frac{1}{2}(\Thi-\Tlow).
\end{equation}
Although $\ev{\MSE}$ and $\chi_{\MSE}$ depend on the CAE run, we find
that $\Tc(L)$ estimated from CAE1 and CAE2 are compatible for all $L$.
This is exemplified in the inset of \figref{fig:errAndSusc} (right):
the estimates from CAE1 and CAE2 overlap within uncertainty.
While the peak location is found to be roughly independent of the CAE run,
we found the peak height to depend on the run, a feature that
also can be seen in the inset of \figref{fig:errAndSusc}. 

In \figref{fig:errCorrelation} (left), we show $\chi_{\rm MSE}$
coming from CAE1 for all lattice sizes. We see that as $L$ increases,
the peak shifts closer to $\Tc$, another indication that
the MSE can feel changes of phase, and is qualitatively suggestive
that $\Tc(L)$ will follow the finite-size scaling relation \eqref{eq:FSS}.
The peak heights generically increase with increasing $L$, although this
increase is not strictly monotonic, as would be expected from a true
order parameter.

In \figref{fig:errCorrelation} (right) we plot $\ev{\MSE}$ against $\ev{|m|}$ for CAE1
for all lattice sizes. We see a clear correlation of $1-\ev{\MSE}$ with the
order parameter, further reinforcing our interpretation that $\ev{\MSE}$ is
sensitive to phase changes.  We thus assume that $\ev{\MSE}$ can be viewed
as an order-parameter analogue, at least to the extent that a meaningful
$\Tc(L)$ can be extracted from peaks in $\chi_{\MSE}$. Since $\Tc(L)$
estimated from CAE1 and CAE2 are compatible, we focus the remainder of our
analysis on CAE1.

\begin{figure}
    \centering
    \includegraphics[width=\linewidth]{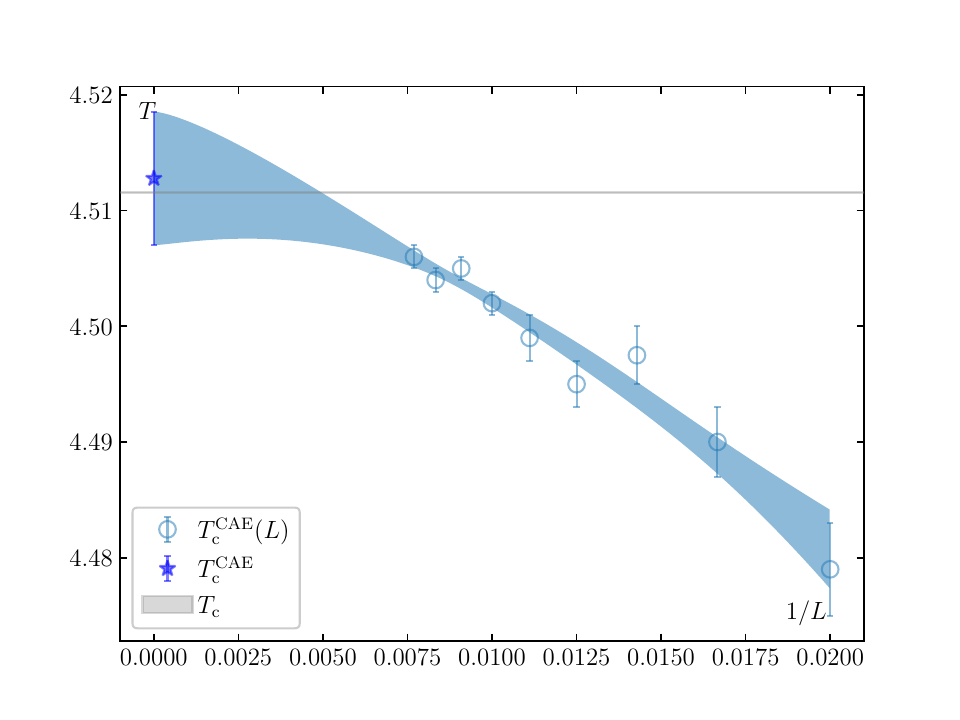}
    \caption{Extraction of $\Tc$ and $\nu$ for the 3D Ising model using the 
    MSE. The horizontal, gray band indicates the literature average for $\Tc$.
    }
    \label{fig:nuTc}
\end{figure}

Using the $\Tc(L)$ extracted from the CAE, we extract $\Tc$ and $\nu$
using the finite-size scaling relation \eqref{eq:FSS}. In particular we
carry out a three-parameter fit of the form
\begin{equation}
\Tc(L)=\Tc+AL^{-1/\nu}.
\end{equation}
The result of the fit is shown in \figref{fig:nuTc}. We find
$\Tc=\TcRes$ and $\nu=\nuRes$ with $\chidof=1.16$. These results
are in excellent agreement with the literature, albeit with
substantially larger uncertainties. In particular the result for
$\nu$ has a relative uncertainty of 43\%.

\begin{table}
\centering
\caption{Critical parameters of ML model compared against
weighted average over literature results~\cite{Pawley:1984et,Barber:1983mp,livet_cluster_1991,ito_monte_1991,blote_finite-size_1993,LANDAU199441,Blote:1995zik,Talapov:1996yh,gupta_critical_1996,Kim:1996zz,Blote:1999vw,Hasenbusch:2010hkh,Hasenbusch:1996eu,Hasenbusch:1998gh,Ballesteros:1998zm,El-Showk:2014dwa,Kos:2016ysd}.}
\label{tab:finalResults}
\begin{ruledtabular}\begin{tabular}{ccc}
& Literature & CAE  \\ 
\hline
$\Tc$        & 4.511535(31) & \TcRes \\
$\nu$        & 0.62999(34)  &  \nuRes  \\
\end{tabular}\end{ruledtabular}
\end{table}

\begin{figure*}
    \centering
    \includegraphics[width=\linewidth]{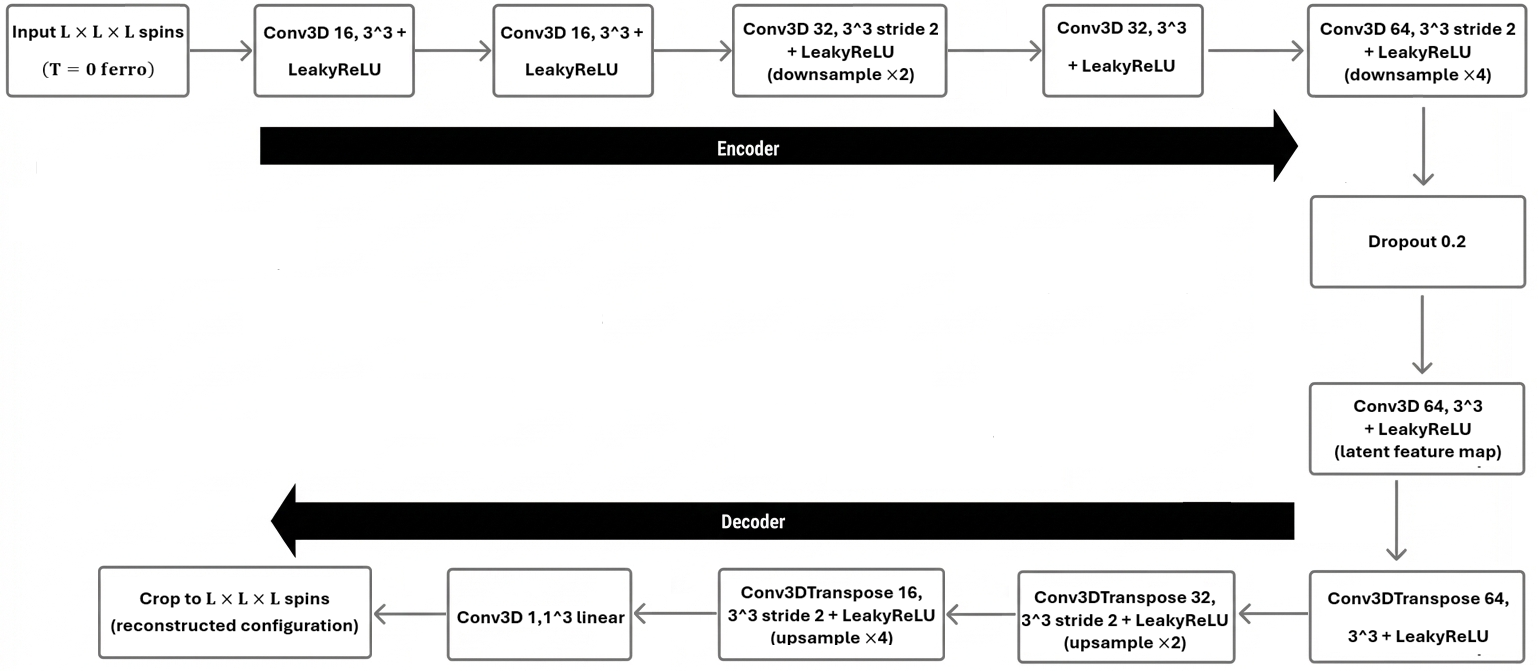}
    \caption{Architecture of the 3D CAE used in this work.}
    \label{fig:Network_Arch}
\end{figure*}

\section{Summary and Outlook}\label{sec:summary}

In this paper, we extended the application of a minimal training
regimen from the 2D Ising to the 3D Ising model and from a supervised binary classifier to 
an unsupervised autoencoder. We find that
the MSE is sensitive to a change of phase and correlates strongly
with the magnetization. From its susceptibility,
we are able to extract $\Tc$ and $\nu$ with good accuracy.
Our results reinforce that one class is sufficient for finding
phase transitions and demonstrate that a one-class autoencoder can recover 
nontrivial critical behavior 
directly from microscopic data.

For both the 2D and 3D Ising models, one- and two-class training regimens are
capable of extracting $\Tc$ with a precision near the per-mille level.
The critical exponent $\nu$ has a large relative uncertainty.
Nevertheless, we take this as convincing evidence that
one-class regimens can be used to accurately and precisely extract
critical temperatures in many statistical physics models,
and they discover enough about the critical behavior of the system
to provide at least rough estimates of critical exponents in
some cases. The impressive ability of minimal training sets to find
$\Tc$, even when the system complexity is increased, increases our confidence 
that this approach will be useful for identifying phase boundaries
for systems with crossovers or systems where the order parameter
is not known {\it a priori}.

\acknowledgments
This material is based upon work supported by the National Science Foundation under grants No. PHY-2208724, PHY-2116686 and PHY-2514763, and within the framework of the MUSES collaboration, under Grant No. OAC-2103680. This material is also based upon work supported by the U.S. Department of Energy, Office of Science, Office of Nuclear Physics, under Award Number DE-SC0022023, as well as by the National Aeronautics and Space Agency (NASA) under Award Number 80NSSC24K0767.

\appendix

\section{CAE architecture}\label{ap:CAE}

The ML model applied in this study is a one-class convolutional autoencoder
written in Python with Tensorflow~\cite{tensorflow} and
Keras~\cite{Keras} libraries. Its
architecture is illustrated in Fig.~\ref{fig:Network_Arch}.
The network consists of a 3D convolutional encoder followed by 
a symmetric decoder. It operates directly on spin configurations of shape 
$L \times L \times L \times 1$.
The encoder progressively reduces the spatial resolution of the input using strided 
3D convolutions. 

%
%

Two convolutional layers with $3\times3\times3$ kernels 
and 16 channels are first applied at full resolution, followed by a stride-2 convolution 
that downsamples the lattice by a factor of two. A further convolution at fixed 
resolution is then applied, after which a second stride-2 convolution reduces the 
spatial resolution by an additional factor of two, producing a latent feature map 
with 64 channels. All convolutional layers use LeakyReLU activations.\footnote{Compared to 
standard ReLU, LeakyReLU maintains a nonzero gradient for negative inputs, which improves 
training stability and preserves sensitivity to both signs of the spin field. This is 
particularly important for autoencoders operating on $\pm 1$ spin configurations and 
for reconstruction-based anomaly detection.}  
A dropout layer with rate 0.2 is applied at the bottleneck to combat overfitting.
The decoder restores the original lattice resolution through a sequence of learned 
3D upsampling layers. Starting from the latent representation, successive 
layers increase the lattice resolution by factors of two until the original lattice size 
is recovered. A final $1\times1\times1$ convolution with linear activation maps the 
reconstructed feature channels at each lattice site to a single spin value. The output 
therefore has spatial size $L\times L\times L$. To ensure consistency of the output 
size, the reconstruction is cropped to $L\times L\times L$ when necessary.
This fully convolutional architecture preserves spatial locality and translational 
structure, making it well-suited for learning the characteristic patterns of the 
ordered phase directly from microscopic configurations.

\bibliography{bibliography}
\end{document}